\documentclass[twocolumn]{aastex63}
\usepackage{graphicx}

\newcommand{\kms}{$\rm km \ s^{-1}$}
\newcommand{\Msun}{$\rm M_\odot$}

\begin{document}

\title{The ``Dark-Matter Dominated" Galaxy Segue~1 Modeled with a Black Hole and no Dark Halo}

\author[0009-0007-4163-9336]{Nathaniel Lujan}
\altaffiliation{These authors contributed equally as co–first authors.}
\affiliation{Department of Physics \& Astronomy, The University of Texas at San Antonio, One UTSA Circle, San Antonio, TX 78249, USA}

\author[0000-0002-8433-8185]{Karl Gebhardt}
\altaffiliation{These authors contributed equally as co–first authors.}
\affiliation{Department of Astronomy, The University of Texas at Austin, 2515 Speedway Boulevard, Austin, TX 78712, USA}
\email{gebhardt@utexas.edu}

\author[0000-0003-3457-7660]{Richard Anantua}
\affiliation{Department of Physics \& Astronomy, The University of Texas at San Antonio, One UTSA Circle, San Antonio, TX 78249, USA}
\affiliation{Physics \& Astronomy Department, Rice University, Houston,
Texas 77005, USA}

\author[0000-0002-0304-5701]{Owen Chase}
\altaffiliation{These authors contributed equally to the class project.}
\affiliation{Department of Astronomy, The University of Texas at Austin, 2515 Speedway Boulevard, Austin, TX 78712, USA}

\author[0000-0002-1998-5677]{Maya H. Debski}
\altaffiliation{These authors contributed equally to the class project.}
\affiliation{Department of Astronomy, The University of Texas at Austin, 2515 Speedway Boulevard, Austin, TX 78712, USA}

\author[0000-0002-9884-9584]{Claire Finley}
\altaffiliation{These authors contributed equally to the class project.}
\affiliation{Department of Astronomy, The University of Texas at Austin, 2515 Speedway Boulevard, Austin, TX 78712, USA}

\author[0000-0002-2503-7083]{Loraine V. Gomez}
\altaffiliation{These authors contributed equally to the class project.}
\affiliation{Department of Physics \& Astronomy, The University of Texas at San Antonio, One UTSA Circle, San Antonio, TX 78249, USA}

\author[0000-0001-8470-7289]{Om Gupta}
\altaffiliation{These authors contributed equally to the class project.}
\affiliation{Department of Astronomy, The University of Texas at Austin, 2515 Speedway Boulevard, Austin, TX 78712, USA}

\author[0009-0005-6226-8226]{Alex J. Lawson}
\altaffiliation{These authors contributed equally to the class project.}
\affiliation{Department of Astronomy, The University of Texas at Austin, 2515 Speedway Boulevard, Austin, TX 78712, USA}

\author[0009-0001-1115-4609]{Izabella Marron}
\altaffiliation{These authors contributed equally to the class project.}
\affiliation{Department of Physics \& Astronomy, The University of Texas at San Antonio, One UTSA Circle, San Antonio, TX 78249, USA}

\author[0009-0000-2997-7630]{Zorayda Martinez}
\altaffiliation{These authors contributed equally to the class project.}
\affiliation{Department of Astronomy, The University of Texas at Austin, 2515 Speedway Boulevard, Austin, TX 78712, USA}

\author[0000-0002-3531-4806]{Connor A. Painter}
\altaffiliation{These authors contributed equally to the class project.}
\affiliation{Department of Astronomy, The University of Texas at Austin, 2515 Speedway Boulevard, Austin, TX 78712, USA}

\author[0009-0009-7538-6378]{Yonatan Sklansky}
\altaffiliation{These authors contributed equally to the class project.}
\affiliation{Department of Astronomy, The University of Texas at Austin, 2515 Speedway Boulevard, Austin, TX 78712, USA}

\author[0009-0003-1798-8406]{Hayley West}
\altaffiliation{These authors contributed equally to the class project.}
\affiliation{Department of Physics \& Astronomy, The University of Texas at San Antonio, One UTSA Circle, San Antonio, TX 78249, USA}

\begin{abstract}

The dwarf spheroidal galaxy, Segue~1, is thought to have one of the largest ratios of dark matter to stellar mass. With a reported mass-to-light ratio of over 1000 and being nearby at 23~kpc, Segue~1 is considered one of the best objects to understand dark-matter dominated galaxies. Using orbit-based dynamical models, we model Segue~1. We find the model that fits best requires a black hole mass of
$4\pm 1.5\times 10^5 $~\Msun. The value of the black hole mass is the same with or without a dark halo. The mass-to-light ratio of the stars is poorly constrained by the dynamical modeling, reflecting that Segue~1 is dominated by mass other than stars. Dynamical models that exclude a black hole provide a worse fit and require a dark halo with very small scale radii of around 100~parsecs. Additionally, the zero black hole models require a stellar orbital distribution that is highly radially biased, significantly larger than what has been seen in other systems. The model with a black hole provides an orbital structure that is close to isotropic, more similar to other well-studied systems. We argue that the two-parameter models of stars and black hole provide a better description of Segue~1 than the three-parameter models of stars and two dark halo components. Additional support for a central black hole comes from a significant increase in the central rotation. Using individual velocities, we measure a rotation amplitude of $9.0\pm2.4$~\kms; with a dispersion of 4 \kms\ makes Segue~1 a rotational dominated system in the central region, where a central mass is a viable model to sustain such a rotaton. Segue~1 is likely being tidally stripped at large radii, and we might be witnessing the remnant nucleus of a more massive system that now is dominated by a black hole and a loosely-bound stellar nucleus. Alternatively, given the high black hole mass relative to the stellar mass, Segue 1 is analogous to Little Red Dots in the early Universe.

\end{abstract}


\section{Introduction}


Dark matter and black holes play critical— and often overlapping— roles in galactic dynamics. Though neither directly radiates electromagnetically, they both can determine the motions of bodies that emit light in the vicinity of their strong gravitational potentials. This work illustrates a dynamical modeling approach for disentangling these two influences in a subtle case, the globular cluster Segue 1.

A substantial part of the mass-energy budget of the observable universe is a gravitationally attractive component-- dark matter-- that does not interact with other fundamental particles in the Standard Model, yet is thought to play a dominant role in the formation of galactic structure \citep{Press1974,Navarro1996}. Dark matter, roughly 6 times as abundant as its baryonic matter counterpart, is theorized to undergo self-interactions within the dark sector \citep{Spergel2000}, and may even convert from dark to visible sector and back via high energy processes such as photon-paraphoton kinetic mixing in astrophysical contexts \citep{Anantua2010}. The detailed nature of dark matter, whether sub-eV axion particles \citep{Peccei2008} or asteroid-mass primoridal black hole \citep{Curd2024} or yet larger massive compact halo objects (MACHO) \citep{Alcock2000} is an active research area. Though cosmological simulations of large scale structure formation are agnostic to the sub-grid nature of dark matter, dynamical modeling of relatively small systems, such as the dwarf spheroidal galaxies may reveal stark observational 
signatures of dark halos and the nature of the dark matter particle \citep{bullock17}. 

Another part of the mass-energy budget on the observable universe is black holes, with both strong support from theory and observations. A robust prediction of Einstein's General Theory of Relativity is that spacetime curvature bends the geodesic path of light, with sufficiently compact mass distributions trapping light behind horizons bounding dark, maximally dense regions, 
i.e., 
black holes. For a non-rotating mass distribution, a black hole of mass $M_\mathrm{BH}$ is formed when the mass is contained within its Schwarszchild horizon $r_S=\frac{2GM_\mathrm{BH}}{c^2}\equiv 2M$ (for rotating black holes the outer horizon lies between $M$ and $2M$). Given that the black hole radius scales linearly with its mass, the average density contained within the horizon scales as $1/M_\mathrm{BH}^2$-- and easy requirement for the vast majority of galaxies to satisfy in order to contain central supermassive black holes. Our trillion solar mass Milky Way is an average sized galaxy that hosts a 4 million $M_\odot$ supermassive black hole, Sagittarius A*, at the Galactic Center \citep{ghez08,gillessen09,GRAVITY2022}. Though compact, the central black hole has an outsized impact on galactic dynamics, with the mass tightly related to galaxy properties, e.g., stellar velocity dispersion via the $M-\sigma$ relation $M_\mathrm{BH}\propto \sigma^{4}$  \citep{Gebhardt00, Ferrarese2000} and serving as a central engine governing AGN feedback. For dwarf galaxies, or globular clusters containing thousands to millions of stars, black holes from stellar remnants may partake in multi-body dynamical interactions either accreting or ejecting most of the black holes. There are a handful of cases with black holes in these low-mass systems \citep{gebhardt02,lutzgendorf13,bustamante21}, although some of the results remain controversial.

The dwarf spheroidal galaxy Segue 1 was one of the Milky Way satellite galaxies discovered by the Sloan Digital Sky Survey in the optical+IR \citep{Belokurov2007} as part of the Sloan Extension for Galactic Understanding and Exploration \citep{Newberg2003}. \cite{Simon11} provides stellar radial velocities and show that Segue~1 is dominated by mass other than stars. They infer a dynamical mass that is over 1000 times larger than the light contributed by the stars. Being the most 
dark-component 
dominated galaxy, Segue~1 becomes an essential object to understand the properties of dark matter halos, since the dark matter halo should suffer limited effects from baryonic processes. 

Our approach in this paper is to model Segue~1 with dynamical models that have limited assumptions. It is important to consider mass components that we know exist in the universe based on observations, like stars and black holes, in addition to components that are strongly understood to exist, like dark matter.

We use a distance to Segue~1 of 23~kpc.

\begin{figure}
\includegraphics[width=230pt]{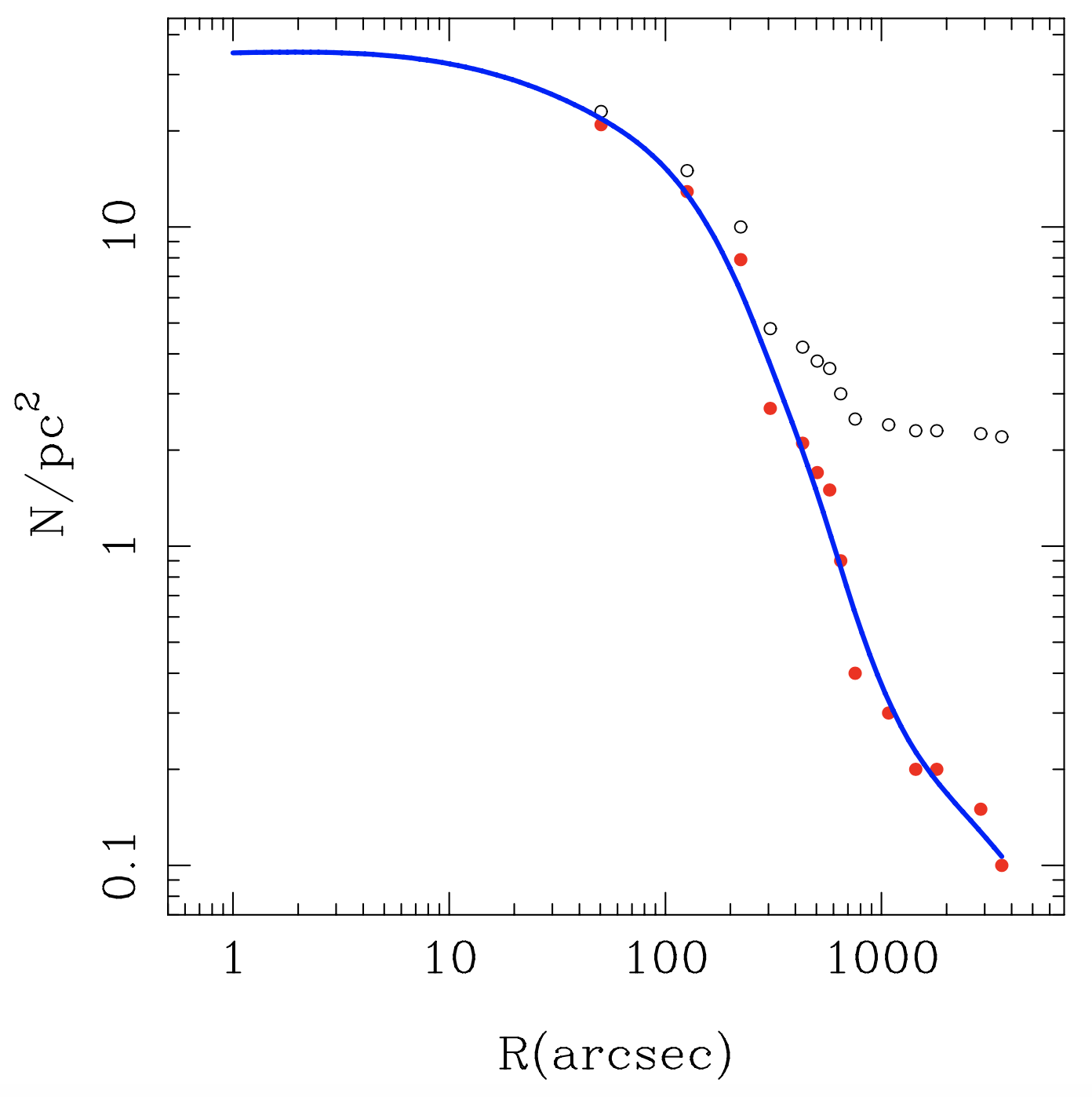} 
\caption{The projected number count profile for Segue~1. The open circles are the data from \cite{Niederste09}. The solid red circles are after subtraction of the tidal effects. The blue line is the smoothing spline that we use for the deprojection.
}
\label{fig:numct}
\end{figure}

\section{Observational Properties of Segue~1}

Typically for dynamical models of diffuse sources, one would measure a surface brightness profile representing the distribution of the tracer light. In the case of Segue~1, the object is so sparse on the sky, that we must rely on number counts. \cite{Niederste09} provide stellar number counts. To separate the Segue~1 stars from the galactic stars, they rely on colors, and the low metallicty of Segue~1 helps to discriminate. \cite{Niederste09} provide number counts starting around one arcminute and going out to nine arcminutes.

A further complication is that Segue~1 is being tidally stripped by the Galaxy. The stars in the tidal streams are not dynamically bound to Segue~1 and should be removed from the dynamical analysis, and hence the number density profile. We do this removal by assuming a uniform density of the tidal stream over the full radial range of Segue~1. We take the outer radial value for the density and subtract that from the counts at smaller radii.

In addition to the larger radii correction to the number counts from the tidal effects, we have to extrapolate the number density into the central regions. This extrapolation is required for the dynamical models in order to understand the tracer population distribution in the center. The amount of mass in stars within the extrapolated region is small enough that it will have an insignificant effect on the mass profile.

Figure~\ref{fig:numct} shows the estimated projected number density profile using the data from \cite{Niederste09}. The open symbols are the observed values, the red solid points are after subtraction of the tidal feature, and the line is our smoothed estimate where we show the extrapolation into the central region. We follow the same method of smoothing and deprojection of the number density profile as outlined in \cite{gebhardt96}, where they provide a non-parametric estimate of the 3d density profile given the projected profile.

\begin{figure}[h]
\includegraphics[width=230pt]{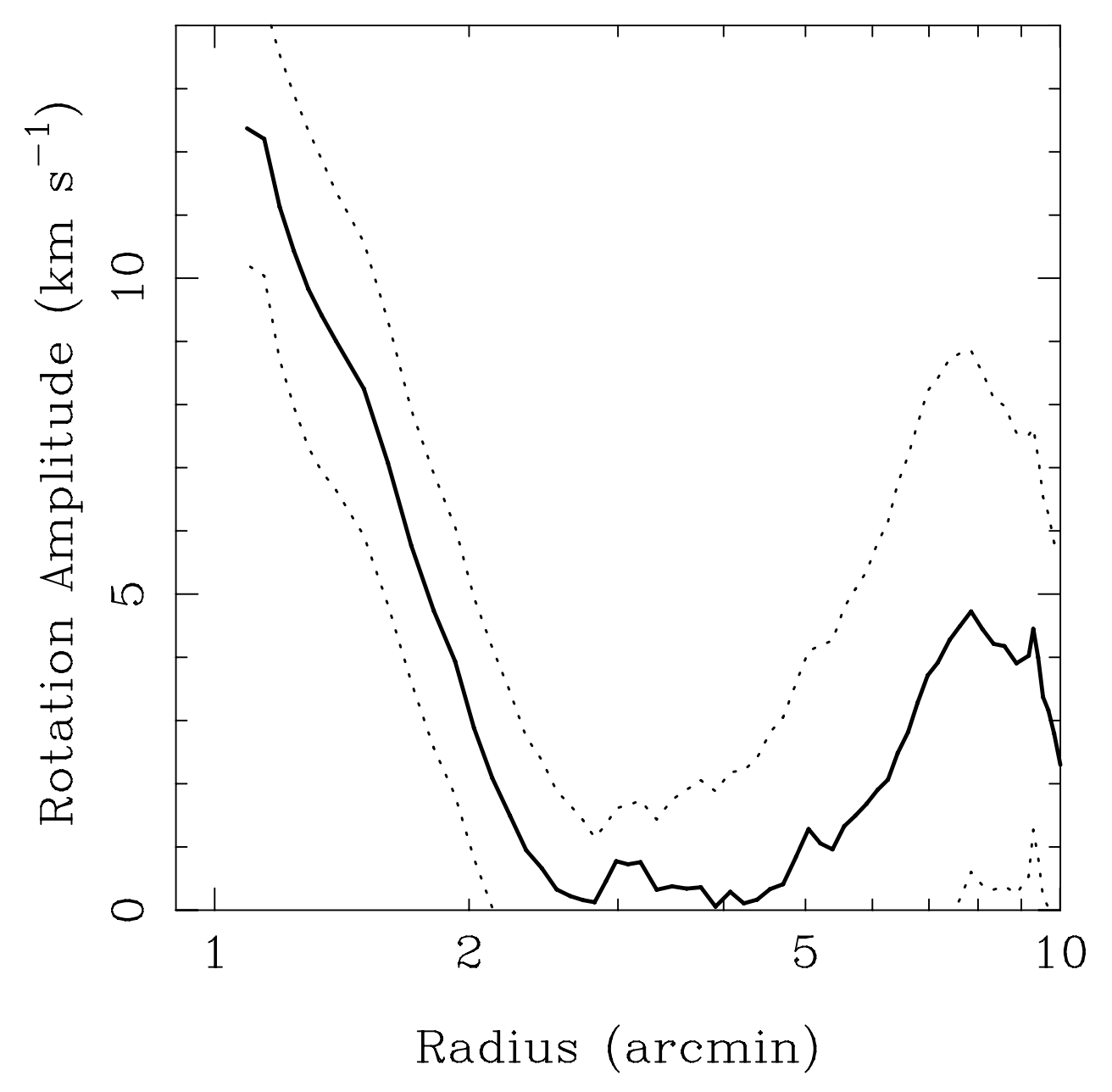} 
\caption{Amplitude of the rotation as a function of radial bin. We use radial bins of 21 stars for most of Segue~1, and going to a minimum of 15 stars at center and outer bins. In each radial bin, we fit a sinusoid as a function of position angle, and plot the amplitude of the fit. The thick solid line represents that amplitude, and the dotted line represent the 68\% confidence bands. There is a large rise in the rotation in the central region, significant at over 99\%. Beyond 2\arcmin\ there is no significant rotation.
}
\label{fig:rot}
\end{figure}


\begin{figure*}
\includegraphics[width=510pt]{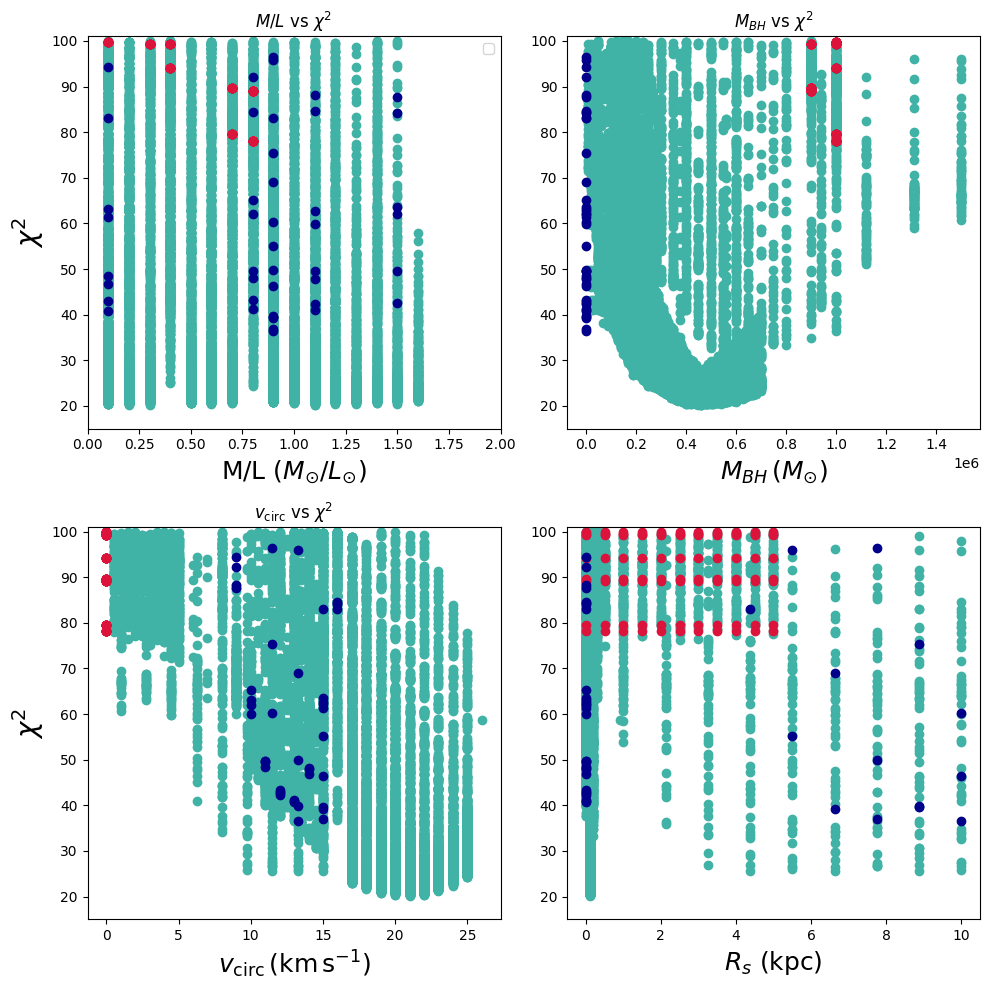} 
\caption{$\chi^2$ versus each of the four parameters. The top left is stellar mass-to-light ratio, the top right is black hole mass in solar masses, the bottom left is dark matter circular velocity, and bottom right is dark matter scale radius. In each panel, we plot all models and a contour along the bottom in $\chi^2$ determines the best fit value and the uncertainty. The red points highlight those models that have no dark halo, and the blue points highlight those models that have no black hole. The green points are the other models that vary all four parameters.
}
\label{fig:chi2}
\end{figure*}

Stellar kinematics come from \cite{Simon11}. In that paper, they carefully select members of Segue~1 using their spectra taken with high resolving power, which they also use for the radial velocity measurements. They bin the velocity into 5 radial bins and measure the velocity dispersion. We generate a line-of-sight velocity distribution from the velocity dispersion data. Since they only provide a velocity dispersion and no higher-order moments, we assume a Gaussian profile. A proper fit would be to use the individual velocities to generate  a binned line-of-sight distribution. The small number of stars in this case makes the assumption of a Gaussian profile a simple solution.

\subsection{Central Rotation}

Segue~1 demonstrates a large rise in the rotation in the central region. Figure~\ref{fig:rot} shows the rotation amplitude as a function of radial bins. We use a similar analysis as in \cite{gebhardt1995}, where we fit a sinusoid to the individual velocities in an annulus. For the majority of radial bins, we use 21 individual stars for the fit, and we go down to 15 stars at the central outer edges. The amplitude of the sinusoid fit is the rotation velocity. We use a least-squared minimazation for the fit and a Monte Carlo for the 68\% uncertainties. The thick black line in Fig~\ref{fig:rot} is the velocity amplitude and the dotted lines represent the 68\% uncertainties. The velocity amplitude rises to over 10\kms\ at 1\arcmin, coming from 15 individual velocities. The smallest annulus where we have 21 stars is at 1.4\arcmin. That annulus provides a rotation amplitude of $9.0\pm2.4$~\kms, with a significance of rotation above 99\%. Going down to smaller radial bins with number of stars less than 15 shows a continual increase in rotation up to 14\kms.

\cite{Simon11} measure a velocity dispersion of Segue~1 of 4\kms. A rotational amplitude of 9--14\kms\ in the central region is a dramatic increase. The two main options for generating such a large increase is either 1) a very large rise in the central density or 2) Segue~1 is being tidally destroyed all the way into the center.

\section{Dynamical Modeling}

We use orbit-based models as originally described by \cite{schwarzschild79}. These models provide the most general solution assuming axisymmetry and dynamical equilibrium. There are multiple examples and tests for these models and they have been heavily used \citep{gebhardt00b,siopis09,vandenbosch12}. The models presented here are the same as used in \cite{gebhardt00b}.

For these models, we assume a gravitational potential, sample the stellar orbital phase space, and then fit orbital weights that provide the best match to the kinematic data. We change the gravitational potential, repeat the process, and find the goodness-of-fit for each potential. We generally run about 20k orbits for each model. The goodness-of-fit is the $\chi^2$ of the model and data. The most important aspect of these models is to make sure phase space is sampled both in range and density of parameters.

In addition to these orbit-based models, other state-of-the-art models are Jeans analysis with various assumptions like JAM \citep{cappellari08} and action integral as applied to the Galaxy \citep{binney23}. We focus on the orbit-based models as they are the most general.

\begin{figure}
\includegraphics[width=240pt]{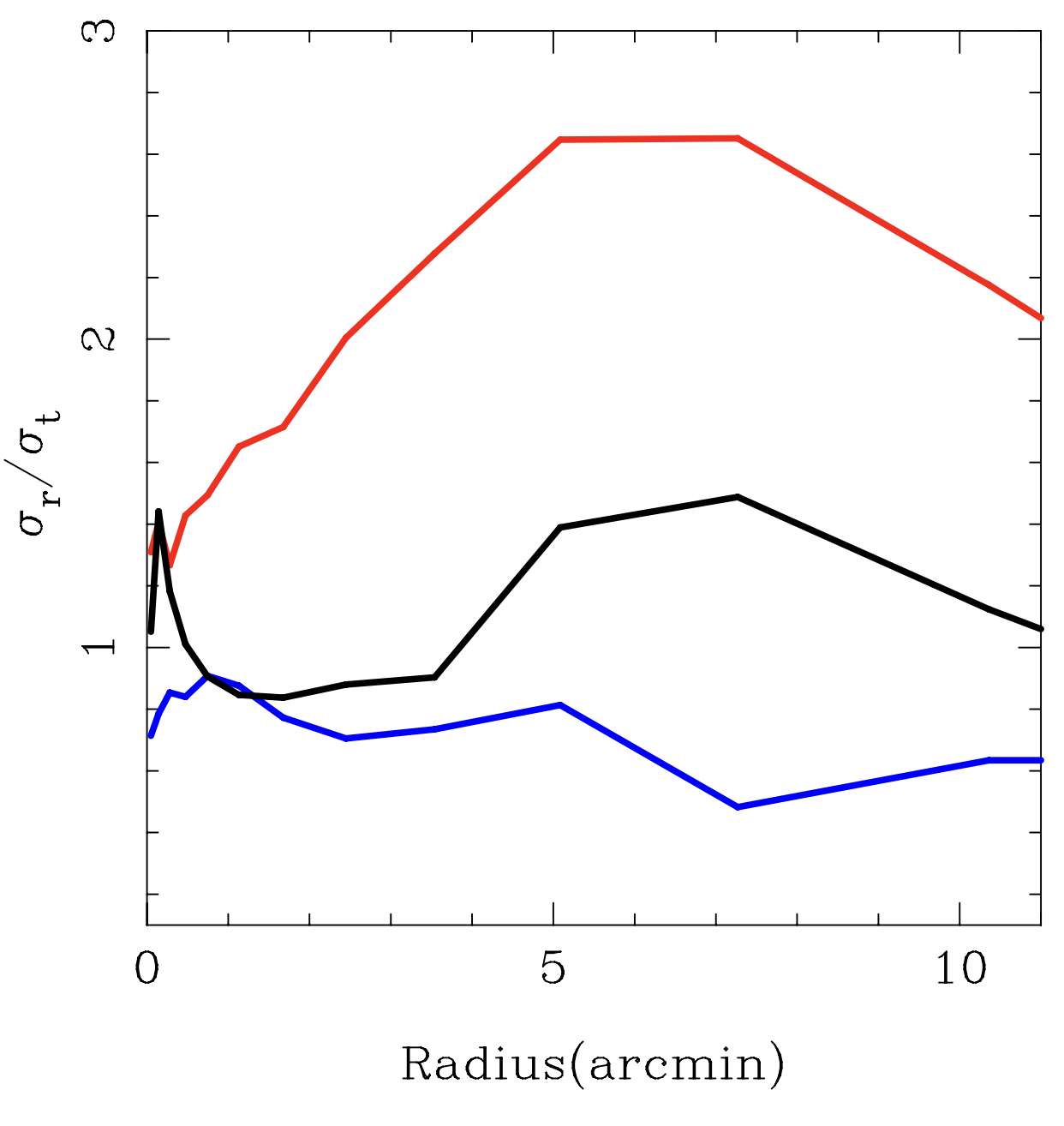} 
\caption{Ratio of internal dispersions $\sigma_r/\sigma_t$ for three different models. The blue and black lines represent models that include a black hole with a mass of approximately $4 \pm 1.5 \times 10^5 \, M_\odot$. These models show close to isotropic dispersions in the central regions of Segue~1. In contrast, the red line represents a model with a dark halo but no black hole, exhibiting significant radial anisotropy. Such high radial anisotropy values are uncommon in other stellar systems.}

\label{fig:sigrt}
\end{figure}

\subsection{Modeling Results}

As inputs to the dynamical models, we require the density distribution of the tracer population, the kinematics for the tracer population, and the profile for a particular gravitational potential. Instead of a non-parametric gravitational potential, we use four parameters: the stellar mass-to-light ratio, black hole mass, dark matter circular velocity, and dark matter scale radius. The dark matter profiles uses a cored logarithmic, which is often used in galaxy models. \cite{bustamante21} finds that the cored logarithmic provides a better fit than a NFW \citep{nfw97} for the dwarf spheroidal galaxy, Leo~1, which is a similar object to Segue~1. 

After defining the gravitational potential, we sample orbits over the available phase space. By forcing the fit to the density profile of the tracer population, we impose consistency between the input stellar profile and the output tracer profile. The fit to the kinematic data provides the one number goodness-of-fit, in terms of $\chi^2$, for each attempted model. Sampling the four parameters, we run about 20,000 models.

Figure~\ref{fig:chi2} shows the $\chi^2$ versus each of the four parameters. The top-left plots the stellar mass-to-light ratio, top-right is the black hole mass in solar units, bottom-left is dark matter halo circular velocity in \kms, and bottom-right is dark halo scale radius in kpc. Each panel plots all attempted models, which is why the points cover much of the region. In order to provide the best fit value and uncertainty, one would use a trace along the bottom of the $\chi^2$ contour.

In this Figure, we highlight models where there is no dark halo as red points, and those with no black hole as blue points. This distinction is important for interpretation of the results and orbital structure.

\subsection{Models with no black hole}

A three parameter fit of the stellar $M/L$, dark matter circular velocity, and dark matter scale radius provides a adequate fit only when allowing for dark halo scale radius that are small. In that case, the dark matter profile begins to mimics a centrally concentrated object, especially since the inner-most kinematic measure is around 10 parsecs \citep{Simon11}. If we use a dark matter profile that is more typical as to what has been used for these systems, we find a model that has a significantly worse fit.

Having a bias for a dark halo with no black hole, the best fitted model has a scale radius around 70~parsecs with a dark halo circular velocity larger than 5~\kms. We do not constrain the upper limit to the circular velocity for the cored-logarithmic profile.

\subsection{Models with a black hole}

A two parameter fit of a stellar mass-to-light ratio and a black hole mass provides a adequate fit to the kinematics. This model is, in fact, better than the three parameter model with a dark halo and stars. When including a dark halo, the fit obviously improves, with the mass of the black hole that best fits basically unchanged. While the dark halo properties are basically unconstrained, the black hole mass is well determined. Thus, this is a case where a two-parameter fit actually fits better than the three-parameter fit, implying the parameterization of the dark halo is not ideal (or there is no need for a dark halo).

We try models with both a black hole and dark halo, still varying the stellar mass-to-light. These models are shown as the green points in Figure~\ref{fig:chi2}. In the top-left panel, one notices that the stellar mass-to-light ratio is minimized at zero. The implications are that the stars contribute nothing to the gravitational potential. We studied a high density sampling in the black hole mass around a value of zero for the stellar mass, which can be seen as the extensions for dark matter circular velocity of 15~\kms\ and scale radius of 0.5~kpc. This high density sampling does not change the best fit black hole mass. We do not explore these models further since we know that stars have mass, and we should at least include a contribution of the stars to the gravitational potential.

Using the bottom contours for the black hole mass $\chi^2$ plot for either the red or the green points provide about the same values of $4\pm 1.5\times 10^5 $~\Msun.

\subsection{Stellar Orbital Structure}

The orbit-based models allow for the most general distribution of dispersion anisotropies. That is, we do not force any parameterization on the internal dispersions. Not only does this allow for the best fit to the observations, but also allows us to explore whether the orbital structure is consistent with other systems. 

Figure~\ref{fig:sigrt} shows the ratio of the radial to tangential dispersions for three representative models. For each model, we average over all angles, and plot as a function of radius. For most stellar systems, the trend is typically to have close to isotropic orbits with an increase in the tangential dispersion closer to the black hole \citep{gebhardt11,thomas14}. The increase in tangential anisotropy is thought to be due to adiabatic destruction of stars in radial orbits that get close to the black hole \citep{milos01}.

The three models in Figure~\ref{fig:sigrt} are no black hole in red, no dark halo in blue, and best overall (black hole and dark halo) in black. The no dark halo and the best overall model are very similar, and show close to isotropic orbits with an increase in the tangential component in the central regions. The no black hole model requires significant radial anisotropy. This high level of radial anisotropy has not been seen in other systems \citep{gebhardt03} and is likely improbable.

{\subsection{Dark Matter Profile}
Early interpretations of Segue 1 indicate a significantly more massive and concentrated dark matter halo. This perspective was primarily based on the Navarro-Frenk-White (NFW) cuspy profile, which resulted in an extremely high mass-to-light ratio. However, upon comparing the cored dark matter profile with the NFW cuspy profile, we found that the cored dark matter profile provides a significantly better fit to the data, yielding a chi-squared value of 20, as opposed to a value of 24 for the NFW profile. Both models suggest the existence of a central supermassive black hole (SMBH) with a mass of $4 \pm 1.5 \times 10^5 
\text{M}_\odot$.

The better fit of the cored profile is consistent with previous studies that demonstrate core-like profiles produce flatter rotation curves, which align more closely with observations of ultra-faint dwarf galaxies \citep{gentile2004}. These flatter profiles are also more vulnerable to tidal disruption during the formation of larger galaxies due to their reduced gravitational binding. This observation supports research indicating that dwarf galaxies without central cusps are more susceptible to disruption, resulting in inhibited or truncated star formation \citep{mashchenko2005}. Furthermore, the observed variation in metallicity within Segue 1—where certain stars exhibit extremely low metallicities ([Fe/H] $\sim$ -3.7) while others display higher values ([Fe/H] $\sim$ -1.5)—reinforces the idea of intermittent episodes of star formation \citep{frebel2014, kirby2013}. Simulations in \citep{penarrubia2010} have demonstrated the vulnerability of cored dark matter halos to tidal forces, showing that Milkey Way satellite galaxies embedded in core dark matter halos are easily stripped during interactions with the host galaxy's baryonic disk \citep{penarrubia2010}. This susceptibility could contribute to the faint nature of Segue 1, as tidal stripping diminishes its baryonic and dark matter, thereby limiting its capacity to sustain star formation.

\subsection{Concerns}

The small amount of kinematic data is clearly a concern, implying a possibility of over-interpretation of the results. While this over fitting needs to be considered, the more robust approach is to use orbit-based models in order to understand the possible parameter space. In this case, the model with two parameters for the gravitational potential provide a better fit than those with three parameters.

Another concern is that the central kinematic measurement is biased high due to stellar binaries inflating the projected velocity dispersion. This concern is studied in \cite{Simon11} where they estimate an upper limit of 1-2~km/s that might need to be removed in quadrature from the measured dispersion. They infer a binary fraction around 10\%, consistent with studies in globular clusters \citep{gebhardt94}. Thus, it is unlikely that binaries are the cause for the increase in the velocity dispersion in the central regions.

Segue~1 is being tidally-stripped by the Milky Way. We had to model the tidal effects in the number counts, similar to what was done in \cite{bustamante21}. Any correction to the tidal stripping would affect both the black hole models and the dark halo models. The central region, where the tidal effects are minimal, drive the fits to the black hole models, so the tidal model is unlikely to be significant for the black hole mass estimate. \cite{bustamante21} showed a similar lack of effect for Leo~1.

\section{Implications}

The work in \cite{hayashi23} models Segue~1 using an NFW halo and find a steep central density slope, with the highest density yet reported. The central density that we infer is similar to that of \cite{hayashi23}, except our central density is due to a black hole as opposed to highly-concentrated dark matter. The alternative explanation of a black hole is a more natural explanation than the steep dark matter profile.

If Segue~1 is an accreted galaxy, as suspected, it should not be a surprise that there is a black hole in the center. Except for one galaxy, M33 \citep{gebhardtM33}, we have yet to discover a galaxy that has either no black hole or a significantly low upper limit.

The ``dark-matter" dominated galaxy Segue~1 has most of its mass in material that is dark. In this case, that material is primarily in the black hole as opposed to a standard dark halo.

There are two main considerations that may influence our interpretation. First,
the extraordinary nature of a system with 
order $10^5$ solar masses of black hole but only around 
 $3\cdot10^4$ solar masses 
in stars may push models beyond their current regime of applicability.
However, this 
large ratio 
may 
simply reflect the 
extraordinary intrinsic 
nature 
of the 
source. 
The second consideration stems from tidal effects. If the radial anisotropy of Segue 1 is 
because it is being disrupted from equilibrium due to tidal effects, 
then any dynamical modeling is suspect--
and Segue 1 
may indeed be comprised primarily of dark matter and stars originating from a larger population that has been tidally stripped. 
Tidal effects, though, are 
mitigated in our analysis (cf. Fig. 1) which  only focus on the compact central region where tidal effects are minimal.

Our result may reveal a consequence of the outlier properties of Segue 1, which in addition to its extremely high mass-to-light ratio,  also has a very wide metallicity distribution (with most stars $\Delta[Fe/H]>0.8$ dex \citep{Webster2016}). This lends support to a brief, clustered star formation scenario early in the history of Segue 1 in which it never amassed a large stellar population to be stripped to begin with.Similar structures, with very massive black holes and low stellar mass, are seen in the early Universe \citep{taylor2025caperslrdz9gasenshroudedlittle, matthee2024environmentalevidenceoverlymassive}-- typically called Little Red Dots. We may be witnessing a larger phenomenon of local dark remnants (with no dark matter halo). In the future it may be illuminating to carry out a similar investigation with 
other dwarf spheroidals.}

\acknowledgments

We are grateful to the University of Texas at Austin and the University of Texas at San Antonio for allowing KG and RA to teach a graduate class in gravitational dynamics. This research is a result of the final project in the classes that were taught in tandem. The project was simply to model Segue~1 using orbit-based modeling, and the final result from the class is important enough to warrant publication. We especially thank the Texas Advanced Computing Center for providing the excellent resources for running all of the dynamical models. This work was supported by a
grant from the Simons Foundation (00001470, RA, NL, and HW).
RA was supported by the Oak Ridge Associated Universities Powe Award at the outset of this investigation.



\end{document}